\documentclass{article}

\usepackage[T1]{fontenc}
\usepackage[english]{babel}
\usepackage{color}

\usepackage[noblocks]{authblk}

\usepackage{graphicx}
\usepackage{float}		

\usepackage{times}

\usepackage{lineno}

\usepackage{amsmath}

\usepackage{upgreek}

\usepackage{setspace}

\usepackage[maxfloats=25]{morefloats}

\usepackage[backend=biber,style=chem-rsc]{biblatex}  
\begin{document}

\title{ Urea-Mediated Anomalous Diffusion in Supported Lipid Bilayers}

\author[1]{E. E. Weatherill}
\author[1,2]{H. L. E. Coker}
\author[1,3]{M. R. Cheetham}
\author[1]{M. I. Wallace\footnote{Corresponding author: mark.wallace@kcl.ac.uk}}
\affil[1]{{\itshape Department of Chemistry, Britannia House, King's College London, 7 Trinity Street, London, SE1 1DB, United Kingdom}}
\affil[2]{{\itshape Chemistry Research Laboratory, Department of Chemistry, University of Oxford, Oxford OX1 3TA, United Kingdom}}
\affil[3]{{\itshape Cavendish Laboratory, Department of Physics, NanoPhotonics Centre, University of Cambridge, Cambridge, CB3 0HE, United Kingdom}}
\maketitle
\clearpage

\begin{abstract}
Diffusion in biological membranes is seldom simply Brownian motion; instead, the rate of diffusion is dependent on the timescale of observation and so is often described as anomalous. In order to help better understand this phenomenon, model systems are needed where the anomalous subdiffusion of the lipid bilayer can be tuned and quantified. We recently demonstrated one such model by controlling the excluded area fraction in supported lipid bilayers (SLBs) through the incorporation of lipids derivatised with polyethylene glycol. Here we extend this work, using urea to induce anomalous subdiffusion in SLBs. By tuning incubation time and urea concentration, we produce DCPC bilayers that exhibit anomalous behaviour on the same scale observed in biological membranes.
\end{abstract}

\section*{Key Words}
Anomalous, diffusion, lipid bilayers, membranes, urea.

\section*{Introduction}
Diffusion is a vital process that underpins many cellular functions, including protein organisation \cite{Sheets1997}, signalling \cite{Choquet2003a, Kholodenko2006}, and cell survival \cite{Cheema2011}. In living systems diffusion rarely follows the Brownian motion predicted by a simple random walk model but instead exhibits `anomalous' subdiffusion, whereby the rate of diffusion is dependent on the timescale of observation \cite{Saxton1994}. Anomalous subdiffusion has been observed in 3D in the cytosol \cite{Regner2013} and in 2D in plasma membranes \cite{Hofling2013, Fujiwara2002a, Golan2017a}. The underlying mechanism for anomalous subdiffusion in membranes is thought to involve molecular crowding \cite{Kusumi2005a}, with contributions from slower-moving obstacles \cite{Saxton1987, Berry2014}, pinning sites, and compartmentalisation \cite{Kusumi2005a, Fujiwara2002a, Murase2004}; reviewed comprehensively elsewhere \cite{Saxton2012}. The notion that the cell membrane is a homogenous entity in which lipids and proteins are free to diffuse unhindered, as per the `fluid mosaic model' \cite{Singer1972}, has in recent years been re-evaluated to accommodate increased levels of complexity \cite{Kusumi2005a}.

Anomalous diffusion can be modelled by a power law:

\begin{equation} \label{eq:1}
\big \langle \Delta r^2 \big \rangle = 4 \Gamma \Delta t^\alpha,
\end{equation}

\noindent where the conventional diffusion coefficient {\itshape D} is replaced by an anomalous transport coefficient $\Gamma$, whose dimensions change for different degrees of anomalous behaviour.The anomalous coefficient $\alpha$ defines whether the diffusion is normal ($\alpha = 1$), sub-diffusive ($\alpha < 1$) or super-diffusive ($\alpha > 1$).
The units of $\Gamma$ vary with the degree of anomalous behaviour, which presents a challenge of interpretation. However, by de-dimensionalising the observation time \cite{Saxton1994} with a `jump time' $\tau$,

\begin{equation} \label{eq:2}
\big \langle \Delta r^2 \big \rangle = 4D\Delta t \Big( \frac{\Delta t}{\tau} \Big)^{\alpha-1},
\end{equation}
\noindent the length-scale $\lambda$ associated with the 2D anomalous behaviour can be defined ($\lambda = \sqrt{4D\tau}$).

Artificial bilayers have been critical in furthering our understanding of anomalous diffusion \cite{Schutz1997, Ratto2003, Horton2010, Spillane2014a, Wu2016, Rose2015}. In supported lipid bilayers (SLBs),  phase separation \cite{Ratto2003},  protein binding \cite{Horton2010}, and defect formation \cite{Coker2017} have been used to generate anomalous diffusion. Simulations have also played a vital role\cite{Saxton1994, Saxton1989, Saxton2001, Stachura2014, Mardoukhi2015, Koldso2016, Jeon2016, Bakalis2015, Javanainen2013}, in particular those linking the role of mobile and immobile obstacles within the bilayer to the phenomenon \cite{Saxton1987, Berry2014}. Simulations have also provided the means to better interpret single particle tracking (SPT) data \cite{Kepten2015}, as well as methods for discriminating between distinct classes of anomalous diffusion \cite{Metzler2014}.

In order to elucidate the specific molecular mechanisms giving rise to anomalous subdiffusion {\itshape in vivo}, there is a need for experimental models which are able to exhibit readily tuneable anomalous subdiffusion of a biologically relevant magnitude \cite{Saxton2012}. Recently we used SPT to sample anomalous behaviour over four orders of magnitude of time by forming SLBs containing varying mole fractions of lipids functionalised with polyethylene glycol (PEG), thereby controlling nanoscale obstacle formation \cite{Coker2017}. Here, we make use of urea as a chaotropic agent, with reported ability to alter the physical properties of lipid bilayers \cite{Nowacka2012, Costa-Balogh2006, Yu2001, Yeagle1986a}. Urea is present at high concentrations in the tissues of  deep-sea elasmobranchs (sharks, skates, rays) \cite{Smith1929} and is also part of the Natural Moisturising Factor in skin \cite{Rawlings2004}, where it is thought to offer cell membranes protection from osmotic shock due to highly saline or dehydrating conditions by stabilising the lamellar liquid phase. Here we use single-molecule total internal reflection fluorescence (smTIRF) and perform SPT to evaluate urea as a means to induce anomalous diffusion in pre-formed SLBs.

\section*{Materials and Methods}
\subsection*{Materials}
1,2-dicapryl-{\itshape sn}-glycero-3-phosphocholine (DCPC) was purchased from Avanti Polar Lipids (Alabaster, AL). Texas Red 1,2-dihexadecanoyl-{\itshape sn}-glycero-3-phosphoethanolamine triethylammonium salt (TR-DHPE) and 1,2-dipalmitoyl-{\itshape sn}-glycero-3-phosphoethanolamine-N-[methoxy(polyethylene glycol) - 5000] ammonium salt (PEG(5K)-DPPE) was purchased from Lipoid (Ludwigshafen, Germany). Unless stated, all other chemicals were purchased from Sigma-Aldrich. All aqueous solutions were prepared using doubly deionized 18.2 M$\Omega$ cm MilliQ water.

\subsection*{Supported Lipid Bilayers}
SLBs were prepared on glass coverslips by fusion of small unilamellar vesicles (SUVs) \cite{Brian1984} made from 1.77 mM DCPC doped with 1.0 mol\% PEG(5K)-DPPE and $3 \times 10^{-6}$ mol\% TR-DHPE. The addition of PEG-functionalised DPPE (below the mol\% required to induce anomalous diffusion \cite{Coker2017}) helps improve bilayer fluidity by raising the bilayer, thereby reducing interactions between the lipids in the lower leaflet and underlying glass \cite{Albertorio2005}. Texas Red-labelled DHPE was also included in order to assess the diffusive properties of the bilayer using smTIRF.

Lipid mixtures were first dried with nitrogen and placed under vacuum overnight. The dried lipids were hydrated with water and vortexed before tip sonication (Vibracell VCX130PB with CV188 tip, Sonics \& Materials, Newtown, CA)  for 15 minutes at 25\% amplitude. The resulting clear vesicle suspension was centrifuged (3 minutes; 14000 $\times$ \emph{g}) before the supernatant was retained and any titanium residue (from the sonicator probe) was discarded. SUV preparations were stored at 4$^{\circ}$C for up to 48 hours.

Glass coverslips were rigorously cleaned using stepwise bath sonication with DECON-90, MilliQ water, and propan-2-ol for 20 minutes each. Immediately before use, the glass was dried under nitrogen and cleaned with oxygen-plasma treatment for 3 minutes (Diener Electronic, Femto). A well was created on each coverslip using vacuum grease (Dow Corning). The coverslip was heated to 37$^{\circ}$C before 50 $\upmu$L of SUV stock were diluted 1:1 in buffer (250 mM NaCl, 10 mM EDTA, 20 mM Tris pH 7.0) and added to the chamber immediately.  DCPC SLBs were produced by fusion of the SUVs onto the glass coverslip. The vesicles were incubated for 30 minutes before the membranes were washed thoroughly with degassed MilliQ water followed by buffer.

Urea was added (or removed) by buffer exchange via pipetting; all but 50 $\upmu$L of fluid above the SLB was replaced with 200 $\upmu$l of the new buffer (containing 0.2, 0.5, or 1M urea), a minimum of 5 times. Bilayers were imaged 15 seconds after buffer exchange.

\subsection*{Total Internal Reflection Fluorescence Microscopy}

532 nm continuous-wave laser light was focussed at the back aperture of an objective lens (60$\times$ TIRF oil-immersion NA 1.49, Nikon, $\sim$1.4 kW cm$^{-2}$) such that total internal reflection occurred at the coverslip/sample interface. The excited TR-DHPE fluorescence was transmitted through 545 nm dichroic and 550 nm longpass filters before being imaged with an electron-multiplying CCD camera (Andor iXon). The inverted microscope objective was heated to maintain 37$^{\circ}$C at the sample throughout imaging; above the transition temperature for this lipid to ensure the bilayer was in the liquid phase. Bilayers were imaged at an exposure time of 20 ms for 5000 frames.

\subsection*{Single Particle Tracking}
SPT was performed using TrackMate \cite{Tinevez2017a}, a plugin for ImageJ \cite{Schneider2012a}. The space-time co-ordinates of the output tracks were used to calculate mean-squared displacements calculated for different observation times using custom-written procedures in MATLAB (MathWorks) as described previously \cite{Coker2017}.

\section*{Results}
Diffusion of TR-DHPE in the DCPC SLB was fast (6 $\upmu$m$^{2}$ s$^{-1}$) and normal ($\alpha$ = 1.01 $\pm$ 0.01) in the absence of urea (Fig. 1A\&B). In the presence of 1M urea, the diffusion became slower and more anomalous over time (Fig. 1C). $\alpha$ decreased roughly linearly to 0.38 and the transport coefficient ($\Gamma$) showed an approximately exponential decrease to 0.02 $\upmu$m$^{2}$ s$^{-\alpha}$ (Fig1D) over a 10 minute period. Although $\Gamma$ values cannot be directly compared (because they depend on $\alpha$, which is also changing), a linear change in $\alpha$ would be expected to cause an overall exponential change in $\Gamma$, as we report.

\begin{figure}[H]
\label{fig:X}
\centering
\includegraphics[keepaspectratio, width=1\textwidth]{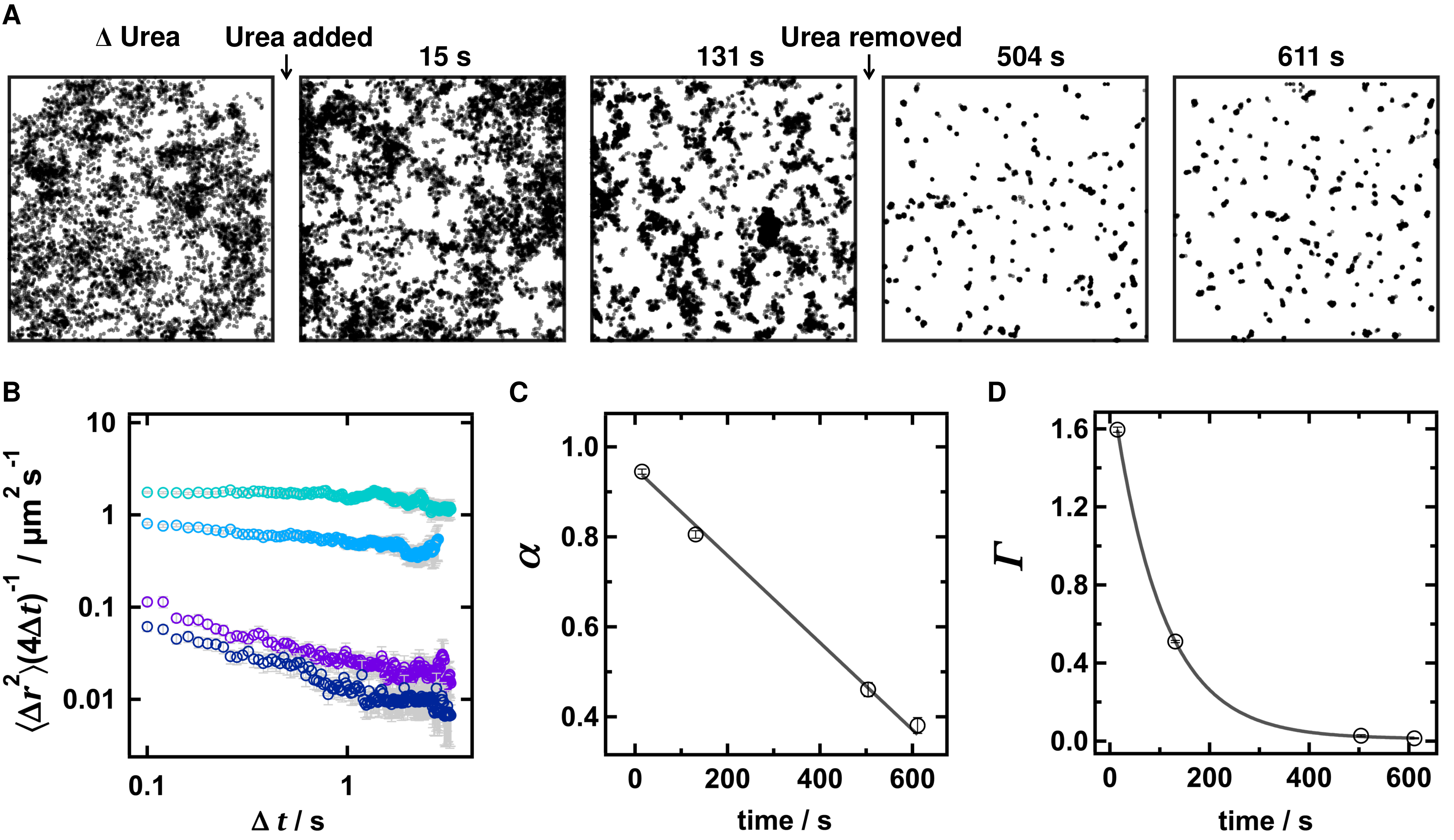}
\caption{\textbf{Time dependence of anomalous behaviour induced by 1 M urea} (A) Spot locations of tracked TR-DHPE in the absence of urea (left) and after the addition of 1M urea at four time points. Urea was removed by buffer exchange at 200-300 s. Image size: 3 $\times$ 3 $\upmu$m (B) Anomalous sub-diffusion increases over time from 15 seconds (turquoise) to 10 minutes (dark blue). (C) Linear decrease of $\alpha$ over time, at a rate of 9.7 $\times$ 10$^{-4}$ s$^{-1}$. (D) Exponential decrease of $\Gamma$ over time, {\itshape t$_{1/2}$} = 69 s. Error bars throughout represent standard errors from a minimum of 250 tracks.}
\end{figure}

Increasing the urea concentration of the buffer surrounding the SLB incrementally from 0 to 1 M, with a fixed short incubation time (15 s), resulted in increasingly slower diffusion (Fig. 2A). The behaviour is largely normal at this short interval, with only a modest decrease of $\alpha$ (to 0.94) at the highest concentration tested (Fig2B). An exponential decrease in $\Gamma$ with increasing urea concentration was observed (Fig. 2C). From the linear relationship between log$_{10}$($\Gamma$/$D$) and $\alpha$ (Fig. 2D) the characteristic length-scale ($\lambda$) associated with the system was calculated to be 45.1 nm, with a jump time ($\tau$) of 86.1 $\upmu$s.

\begin{figure}[H]
\label{fig:Y}
\centering
\includegraphics[keepaspectratio, width=0.7\textwidth]{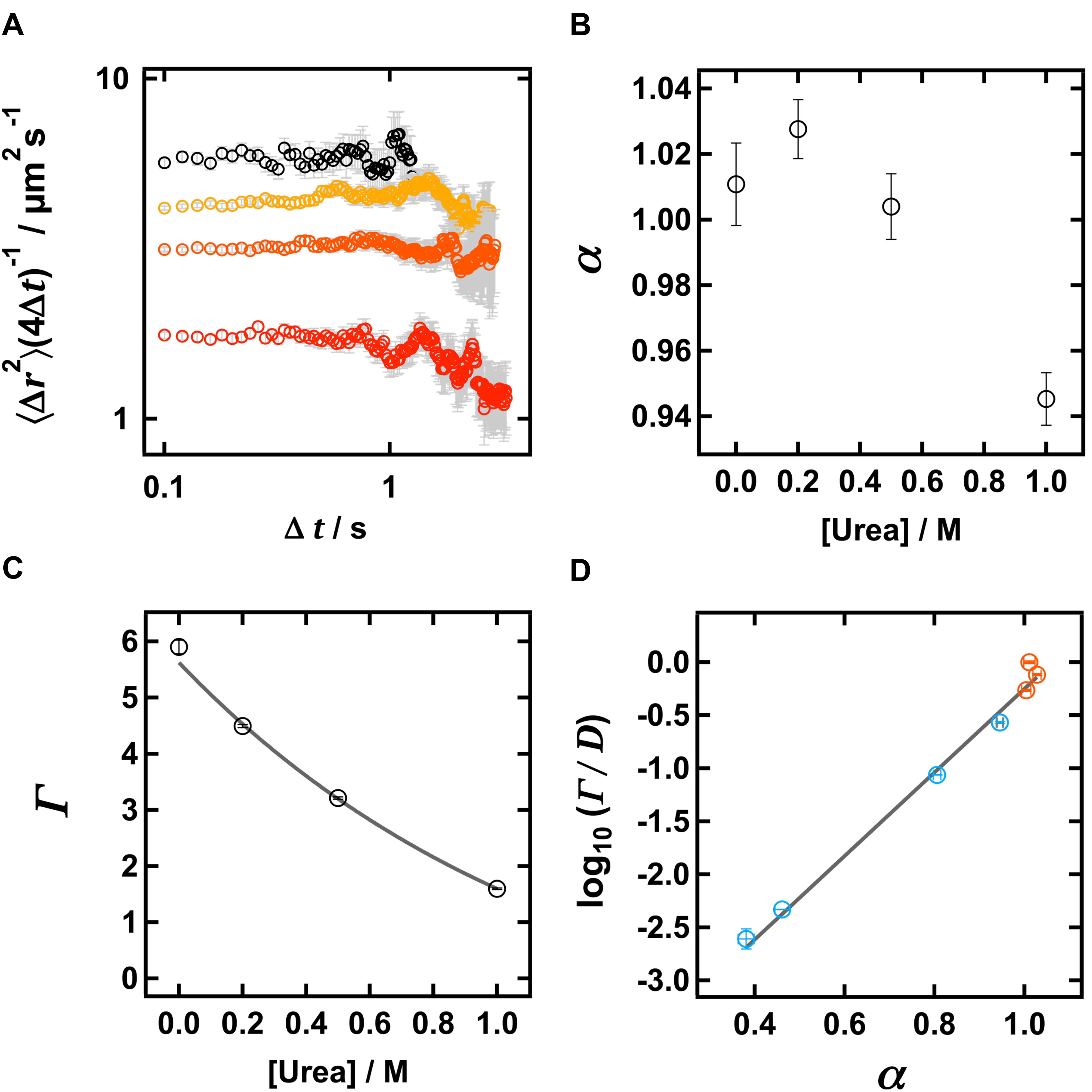}
\caption{\textbf{Effect of urea concentration on lipid diffusion in an SLB} (A) Diffusion of lipids becomes slower as urea concentration of the surrounding buffer is increased from 0 (black) to 1M (red). (B) Decrease of $\alpha$ with increasing urea concentration. (C) Exponential decrease of $\Gamma$ with increasing urea concentration. (D) Plot of log$_{10}$ ($\Gamma$/D) vs. $\alpha$ with linear fit.  Blue: Data from 1M urea timecourse (see Fig. 1); Orange: Data from urea titration (This figure). Error bars represent standard errors.}
\end{figure}

\section*{Discussion}
We observe that urea causes diffusion in DCPC SLBs to become irreversibly slower and more anomalous in a time and concentration-dependent manner. Given our previous experiments reporting defect-mediated anomalous diffusion using PEG-doping of SLBs \cite{Coker2017}, it is appealing to suggest that a similar mechanism must operate for urea. For this case, urea would associate with the bilayer, where its chaotropic nature would act to induce the removal of bilayer patches from the glass coverslip surface, producing defects visible as excluded areas of the surface corresponding to those observed in Fig 1A. However, there is little evidence that urea acts directly to solubilise or otherwise permeabilise lipid bilayers \cite{Costa-Balogh2006}, and this hypothesis would rely on urea acting at the glass-lipid interface.

An alternative explanation for our results would be the action of urea to alter lipid phase behaviour, inducing phase co-existance phases\cite{Nowacka2012}. Unfortunately, the evidence supports a mode of action whereby urea stabilizes the liquid disordered phase \cite{Nowacka2012, Costa-Balogh2006}, suppressing phase separation, rather than encourage it. In our experiments, we observe a decrease in the area fraction of mobile lipids, which is the opposite trend.

A final hypothesis would be the action of urea not on the bilayer, but on the PEG-DHPE. A chaotropic effect on the PEG might act to increase the area fraction occupied by the PEG, which would then again drive the formation of defects in the membrane  \cite{Coker2017}.

The effect that urea has on diffusion appears not only irreversible, but appears to progress even once urea is removed from the bulk solution. The half-life for this process at 1M urea was short (69 s) and was finished after approximately 500 s. We speculate that either our (1000-fold dilution) washing procedure must be ineffective, or there is a more long-lived, direct, interaction between urea and the bilayer. Given the low partition coefficient for urea in lipid bilayers \cite{Diamond1974} and the evidence from studies of multilamellar phases that it remains primarily in the aqueous layers between bilayers \cite{Costa-Balogh2006}, it is difficult to rationalize this as a possible mechanism.

Further work is needed to distinguish between these different possible mechanisms either by viewing the defects directly (e.g. by atomic force microscopy) or by restoring the defects by addition of fresh SUVs.

\section*{Conclusion}
We have presented preliminary findings demonstrating a novel approach to controlling anomalous subdiffusion in SLBs on a scale relevant to biological systems \cite{Murase2004, Schutz1997} by incorporating urea into the aqueous medium surrounding a supported lipid bilayer. Although this work involved the use of DCPC, it would be interesting to extend the method to other, more biologically-relevant lipid compositions. As a complementary method to the inclusion of PEG-lipids, we see potential for this approach for producing a simple membrane model with defined anomaleity,


\section*{Authors' Contributions}
EEW performed the experiments, HLEC and MRC performed the analysis, MIW secured the funding; all authors wrote and reviewed the manuscript.

\section*{Competing Interests}
The authors declare no competing interests.

\section*{Funding}
We thank the European Research Council for providing funding for this work (ERC-2012-StG-106913, CoSMiC).

\section*{References}
\printbibliography[heading=none]
\clearpage

\end{document}